\documentclass[aps,prc,superscriptaddress,showpacs,floatfix,nofootinbib,notitlepage,twocolumn,longbibliography]{revtex4-2}
\usepackage{amsmath,graphicx,float,csquotes,mdframed,appendix,url}
\usepackage[dvipsnames]{xcolor}
\usepackage[colorlinks=true, pdfstartview=FitV, linkcolor=RedOrange, citecolor=Emerald, urlcolor=Cerulean]{hyperref}
\usepackage{amssymb}

\newcommand{\snn}{\sqrt{s_\mathrm{NN}}}
\newcommand{\Pb}{$^{208}$Pb}
\newcommand{\Xe}{$^{129}$Xe}

\graphicspath{{figs/}}

\begin{document}

\title{Probing Nuclear Structure of Heavy Ions at the Large Hadron Collider}

\author{Heikki M\"antysaari}
\affiliation{Department of Physics, University of Jyv\"askyl\"a, P.O. Box 35, 40014 University of Jyv\"askyl\"a, Finland}
\affiliation{Helsinki Institute of Physics, P.O. Box 64, 00014 University of Helsinki, Finland}

\author{Bj\"orn Schenke}
\affiliation{Physics Department, Brookhaven National Laboratory, Upton, NY 11973, USA}

\author{Chun Shen}
\affiliation{Department of Physics and Astronomy, Wayne State University, Detroit, Michigan 48201, USA}

\author{Wenbin Zhao}
\affiliation{Nuclear Science Division, Lawrence Berkeley National Laboratory, Berkeley, California 94720, USA}
\affiliation{Physics Department, University of California, Berkeley, California 94720, USA}

\begin{abstract}
We perform high-statistics simulations to study the impacts of nuclear structure on the ratios of anisotropic flow observables in $^{208}$Pb+$^{208}$Pb and $^{129}$Xe+$^{129}$Xe collisions at the Large Hadron Collider. Even with $40\%$ difference in atomic numbers between $^{208}$Pb and $^{129}$Xe nuclei, the ratios of anisotropic flow in the same centrality class between the two collision systems are strongly affected by the nuclear structure inputs in the initial state. The ratios of $v_2\{4\}/v_2\{2\}$ in these collisions are sensitive to the nuclear skin thickness of the colliding nuclei, providing indirect constraints on the nuclei's neutron skin. Our model predictions serve as a benchmark to compare with experimental measurements.
\end{abstract}

\maketitle

\section{Introduction}

The study of nuclear structure is a cornerstone of nuclear physics, offering profound insights into the emergent properties and interactions of atomic nuclei. Recent \textit{ab initio} approaches aim at describing strongly correlated nuclear systems from solutions of the Schr\"{o}dinger equation with nucleon-nucleon and three-nucleon interactions constructed in an effective theory of low-energy quantum chromodynamics (QCD). These efforts find a natural application in the phenomenology of multi-particle correlations in high-energy nuclear collisions~\cite{Shen:2020mgh, Jia:2021tzt, Bally:2022vgo, Giacalone:2023hwk} and the physics at the future Electron-Ion Collider~\cite{Mantysaari:2023qsq, Mantysaari:2023prg, Mantysaari:2024xmy}. In relativistic heavy-ion collisions, event-by-event snapshots of the colliding nuclei's many-body wavefunctions leave important imprints in the momentum anisotropy of final-state particles because of the ultra-short time duration for the interaction between the two ions at high energy~\cite{Giacalone:2021udy, Achenbach:2023pba, Arslandok:2023utm}.
They are essential inputs for understanding the collective behavior of strongly coupled nuclear matter in the precision era of relativistic heavy-ion collisions~\cite{Rybczynski:2012av, Goldschmidt:2015kpa, Shen:2015qta, Giacalone:2017dud, Giacalone:2019pca, Carzon:2020xwp}. The anisotropic flow measurements at the Large Hadron Collider (LHC) and Relativistic Heavy-Ion Collider (RHIC) have reached sufficient precision to be sensitive to the nuclear structure of colliding nuclei. 
The recent isobar ($^{96}$Ru+$^{96}$Ru and $^{96}$Zr+$^{96}$Zr) collisions at RHIC~\cite{STAR:2021mii} have demonstrated the direct impact of structural properties of nuclei on the collective flow of the produced quark gluon plasma (QGP)~\cite{Hammelmann:2019vwd, Xu:2021uar, Giacalone:2021uhj, Nijs:2021kvn, Zhang:2022fou}. On the experimental side, the ratios of observables between the two isobar systems can cancel most of the systematic errors, which enabled new studies of net baryon and electric charge dynamics at high energies~\cite{Lewis:2022arg, STAR:2024lvy, Pihan:2023dsb, Pihan:2024lxw}.
Therefore, heavy-ion collisions at high energies could serve as a promising tool for imaging the structure of atomic nuclei, providing complementary information to conventional low-energy experimental measurements. 

Constraining the nuclear structure from heavy-ion collisions involves precise measurements in multiple collision systems and sophisticated theoretical models~\cite{Giacalone:2023cet, Ryssens:2023fkv, Giacalone:2024luz,Zhang:2024vkh, Giacalone:2024ixe}. In particular, high-precision simulations play a critical role in this endeavor, enabling quantitative comparisons with the experimental measurements~\cite{Schenke:2020mbo, Xu:2021uar, Hirvonen:2023lqy, Giacalone:2024luz}.
These studies have initiated a new synergy between low-energy \textit{ab initio} theory advancements and high-energy many-body descriptions, from which the nuclear physics community as a whole could benefit.
A future step is to perform robust statistical inference analyses to simultaneously constrain low and high energy model parameters and perform uncertainty quantification on the theoretical models~\cite{Bernhard:2019bmu, JETSCAPE:2020mzn, Nijs:2020ors, Parkkila:2021tqq, Mantysaari:2022ffw, Mantysaari:2022ypp, Heffernan:2023gye, Soeder:2023vdn, Shen:2023awv, Shen:2023pgb, Casuga:2023dcf, Roch:2024xhh, Jahan:2024wpj}.

Although high-energy isobar collisions are ideal systems to study the impact of nuclear structure on heavy-ion observables because final-state effects can be canceled to high precision in the observable ratios, such experimental setups require dedicated planning and resources, which may not be easily accessible. The System for Measuring Overlap with Gas (SMOG2) experiments at LHCb~\cite{LHCb:2021ysy, LHCb:2022qvj, Mariani:2022klh} on the other hand could provide one cost-efficient way to conduct heavy-ion collisions across many nuclear species~\cite{Giacalone:2024ixe}.

In this work, we will explore the impact of the nuclear structure on observable ratios in \Pb+\Pb{} and \Xe+\Xe{} collisions, whose measurements are available at the LHC \cite{ALICE:2024nqd}. Since the mass numbers differ by $\sim$$40\%$ between the two collision systems, we will first identify several observables whose ratios cancel most of the final-state effects based on high-precision simulations with a hydrodynamic + hadronic transport framework. Our model predictions will provide a benchmark for existing~\cite{ALICE:2024nqd} and future experimental measurements.

\section{The simulation framework}

In this work, we perform event-by-event simulations with the state-of-the-art IP-Glasma+MUSIC+UrQMD framework, whose details were described in Ref.~\cite{Schenke:2020mbo}.

In the IP-Glasma initial-state model~\cite{Schenke:2012wb,Schenke:2012hg}, the event-by-event configurations of the \Pb{} and \Xe{} nuclei are implemented as follows. The spatial positions of individual nucleons are sampled with the following deformed Woods-Saxon parametrization,
\begin{align}
    \rho^\mathrm{WS}(r, \theta) = \frac{\rho_0}{1 + \exp\{[r - R(\theta)]/a^\mathrm{WS} \}},
    \label{eq:WoodsSaxon}
\end{align}
with $R(\theta) = R_0^\mathrm{WS} [1 + \beta^\mathrm{WS}_2 Y^0_2(\theta) + \beta^\mathrm{WS}_4 Y^0_4(\theta)$]. Here $Y^m_l(\theta)$ are the spherical harmonics. The values of the Woods-Saxon parameters are listed in Table~\ref{tab:WoodsSaxon}.  We impose a minimum distance $d_\mathrm{min} = 0.9$~fm between individual nucleon pairs, which mimic the repulse interactions between nucleons at short distances~\cite{Broniowski:2007nz}.
When the distance between a nucleon pair is smaller than $d_\mathrm{min}$, we only resample the azimuthal angle of the newly added nucleon until the minimum distance requirement is fulfilled for all nucleon pairs. This algorithm ensures that we still reproduce the desired deformed Woods-Saxon profile in Eq.~\eqref{eq:WoodsSaxon}~\cite{Moreland:2014oya}.
Once nuclear configurations are generated, an independent random 3D rotation is applied to the individual configurations before the collisions.

\begin{table}[tb]
    \centering
    \caption{The values for the Woods-Saxon parameters for $^{208}$Pb and $^{129}$Xe used in the IP-Glasma initial conditions~\cite{Moller:2015fba, Bally:2021qys, Bally:2022rhf}.}
    \begin{tabular}{c|c|c|c|c}
        \hline \hline 
        Nucleus & $R_0^\mathrm{WS}$ (fm) & $a^\mathrm{WS}$ (fm) & $\beta_2^\mathrm{WS}$ & $\beta_4^\mathrm{WS}$ \\ \hline
        \Pb(default) & 6.647 & 0.537 & 0.006 & 0 \\ \hline
        \Xe(1) & 5.601 & 0.492 & 0.207 & -0.003\\ \hline
        \Xe(2) & 5.601 & 0.57 & 0.207 & -0.003\\ \hline
        \Xe(3) & 5.601 & 0.57 & 0.162 & -0.003\\ \hline
        \Xe(4) & 5.601 & 0.57 & 0 & -0.003\\ \hline
        \hline
    \end{tabular}
    \label{tab:WoodsSaxon}
\end{table}

Table~\ref{tab:WoodsSaxon} includes four different sets of Woods-Saxon parameters for the \Xe{} nucleus~\cite{Moller:2015fba, Bally:2021qys, Bally:2022rhf}, with which we will perform high statistics simulations of Xe+Xe collisions. Comparisons of these results will quantify the effects of the nuclear surface thickness parameter $a^\mathrm{WS}$ and elliptical quadrupole deformation $\beta_2^\mathrm{WS}$ on the ratios of flow observables to those in \Pb+\Pb{} collisions. 

In this work, we update the sub-nucleonic structure~\cite{Mantysaari:2020axf} parameters in the model compared to the ones used in Ref.~\cite{Schenke:2020mbo}. 
Here, the parameters are obtained as follows. First, we calculate coherent and incoherent $\mathrm{J}/\psi$ photoproduction at $x_p=2\times10^{-4}$, which is a typical value probed in heavy ion collisions at LHC energies. This calculation is done exactly as in Ref.~\cite{Mantysaari:2022sux}, with the energy dependence obtained by solving the JIMWLK evolution equation~\cite{Mueller:2001uk,Lappi:2012vw}. This setup is constrained by available photoproduction data from HERA experiments and from ultra-peripheral collisions measured at the LHC. Finally, the substructure parameters are extracted from the calculated pseudodata at the smaller $x$ by performing a fit as in Ref.~\cite{Mantysaari:2022ffw}.
The values for the sub-nucleonic parameters are summarized in Table~\ref{tab:subNucleon}.

\begin{table}[h!]
    \centering
    \caption{The values for the sub-nucleonic parameters used in the IP-Glasma initial conditions. The detailed definitions of these parameters can be found in Ref.~\cite{Mantysaari:2022ffw}.}
    \begin{tabular}{c|c|c|c}
        \hline \hline 
        $m$~(GeV) &  $B_G$~(GeV$^{-2}$) & $B_q$~(GeV$^{-2}$) & $\sigma$  \\ \hline
        0.179 &  5.15 & 0.293 & 0.568  \\ \hline
        $Q_s/\mu$ & $d_{q, \mathrm{min}}$~(fm) & $N_q$ \\ \hline
        0.551 & 0.249 & 3 \\ \hline
        \hline
    \end{tabular}
    \label{tab:subNucleon}
\end{table}

The nucleon positions and subnucleon structure determine the thickness function of each nucleus, from which we obtain the color charge densities, which in turn determine the incoming gluon fields via solutions to the Yang-Mills equations \cite{Kovner:1995ts,Krasnitz:1998ns,Krasnitz:2002mn,Lappi:2007ku}. Then we solve for the gluon fields after the collision and evolve them using the source free Yang-Mills equations up to time $\tau_\mathrm{hydro} = 0.4$ fm/$c$ following \cite{Kovner:1995ts,Krasnitz:1998ns,Krasnitz:2002mn,Lappi:2003bi}. The procedure is laid out in detail in \cite{Schenke:2020mbo}.

The system's energy-momentum tensor $T^{\mu\nu}$ is computed from the gluon fields and provides the initial condition for relativistic viscous hydrodynamic simulations performed using MUSIC~\cite{Schenke:2010nt,Paquet:2015lta}. These simulations employ a lattice-QCD based equation of state~\cite{Bazavov:2014pvz, Moreland:2015dvc}. We consider both shear and bulk viscous effects during the hydrodynamic evolution by solving the Denicol-Niemi-Molnar-Rischke (DNMR) theory with spatial gradient terms up to the second order~\cite{Denicol:2012cn}. The fluid cells are converted into hadrons at a constant energy density hyper-surface with $e_\mathrm{sw} = 0.18$~GeV/fm$^3$ using the Cooper-Frye particlization prescription with the Grad's 14-moment viscous corrections~\cite{Huovinen:2012is, Shen:2014vra,Zhao:2022ugy}. These hadrons are then fed into the hadronic transport model (UrQMD) for further scatterings and decay in the dilute hadronic phase~\cite{Bass:1998ca,Bleicher:1999xi}. 

We first calibrate our simulations using \Pb+\Pb{} measurements at 5.02 TeV. Then we vary the nuclear structure parameters for the Xe nucleus as shown in Table~\ref{tab:WoodsSaxon} to study their impact on the ratios of observables between \Xe+\Xe{} and \Pb+\Pb{} collisions.

\begin{figure}[t!]
    \centering
    \includegraphics[width=\linewidth]{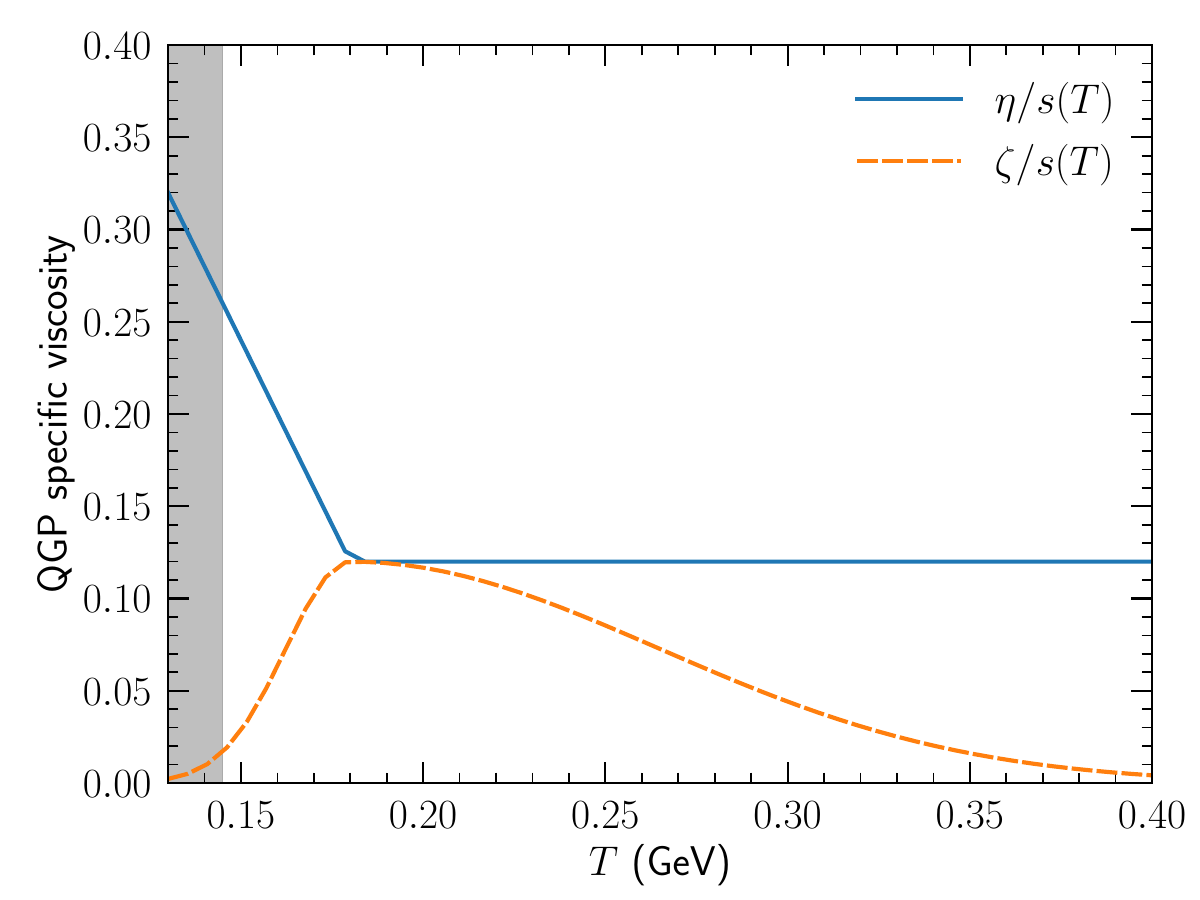}
    \caption{The temperature-dependent QGP specific shear and bulk viscosities used in this work. The grey area indicates the hadronic phase simulated by the UrQMD transport model. }
    \label{fig:QGPviscosity}
\end{figure}

\begin{figure}[ht!]
    \centering
    \includegraphics[width=0.97\linewidth]{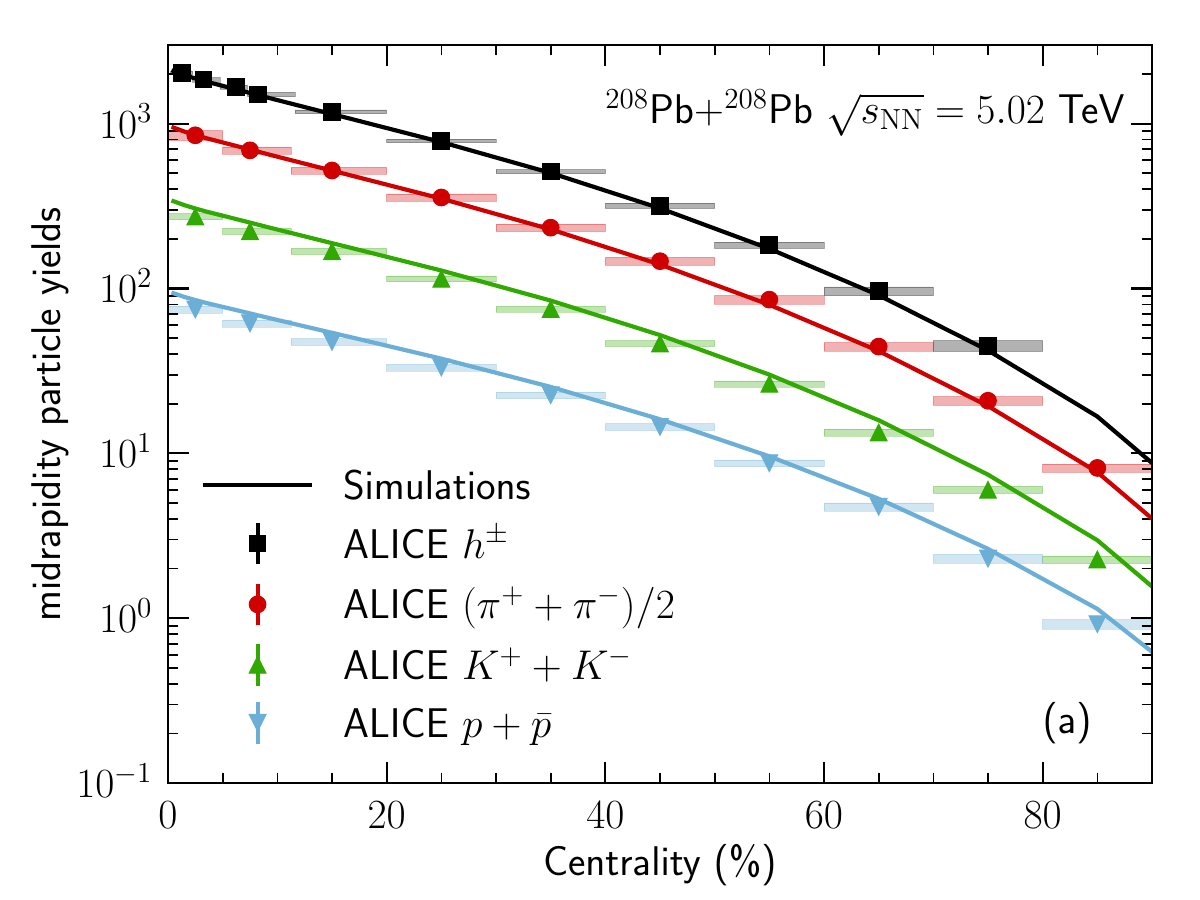}
    \includegraphics[width=0.97\linewidth]{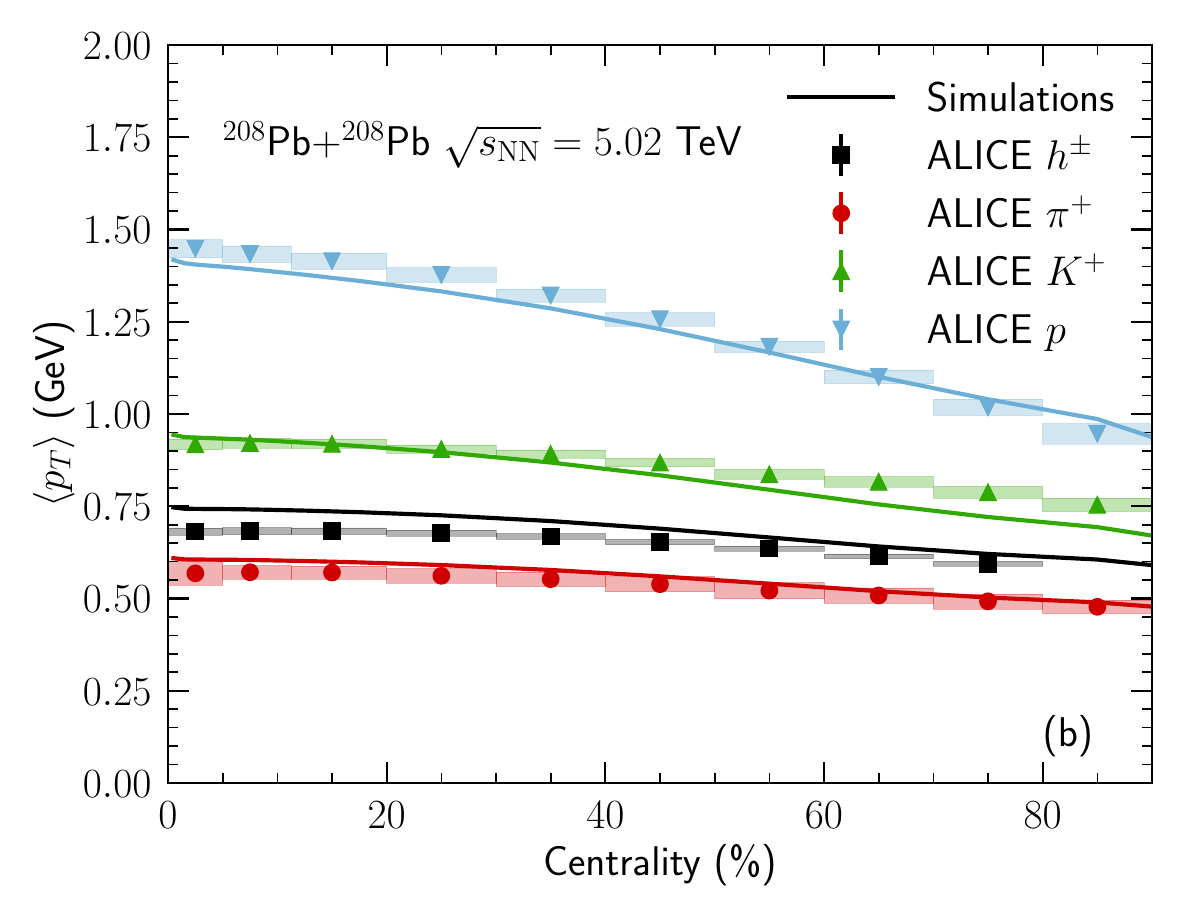}
    \includegraphics[width=0.97\linewidth]{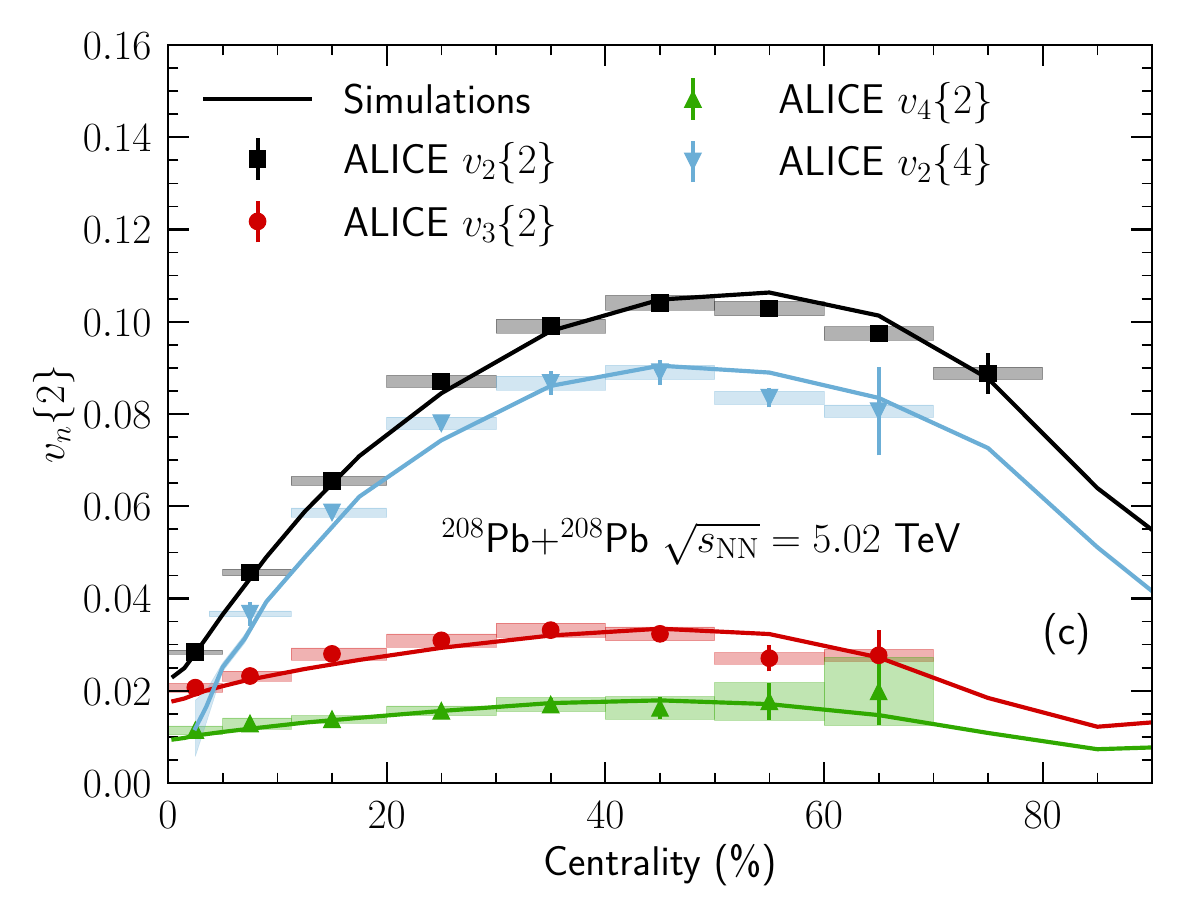}
    \caption{Charged hadron and identified particle yields (Panel (a)) and their averaged transverse momenta (Panel (b)), and the charged hadron anisotropic flow coefficients $v_n\{2\} (n = 2-4)$ and $v_2\{4\}$ are shown in Panel (c) as functions of collision centrality in \Pb+\Pb{} collisions at $\snn = 5.02$ TeV. The model calibration results (solid lines) are compared with the ALICE measurements~\cite{Adam:2015ptt,Acharya:2019yoi,Adam:2016izf,Acharya:2018ihu}. The statistical errors in the calculations are plotted as shaded bands, which are narrower than the line widths for most cases.}
    \label{fig:PbPbobs}
\end{figure}

We introduce the temperature dependence for the QGP specific shear and bulk viscosities shown in Fig.~\ref{fig:QGPviscosity}. The parameterizations are as follows:
\begin{equation}
    \frac{\eta}{s}(T) = \left\{\begin{array}{cc}
       \eta_0 + b(T - T_0) & \quad\mbox{for}\quad T < T_0 \\
       \eta_0  & \quad\mbox{for}\quad T \ge T_0
    \end{array} \right.
\end{equation}
and
\begin{equation}
    \frac{\zeta}{s}(T) = \left\{ \begin{array}{cc}
      \zeta_0 \exp\{-[(T - T_0)/\sigma_+]^2\}  & \,\mbox{for}\, T \ge T_0 \\
      \zeta_0 \exp\{-[(T - T_0)/\sigma_-]^2\}  & \,\mbox{for}\, T < T_0
    \end{array} \right.,
\end{equation}
where $\eta_0 = 0.12$, $T_0 = 0.18$ GeV, $b = -4$ GeV$^{-1}$, $\zeta_0 = 0.12$, $\sigma_+ = 0.12$ GeV, and $\sigma_- = 0.025$ GeV.

Figure~\ref{fig:PbPbobs} shows the model-to-data comparisons with the ALICE measurements of charged hadron and identified particle production, averaged transverse momentum, and charged hadron anisotropic flow coefficients as functions of the collision centrality for \Pb+\Pb{} collisions at 5.02 TeV.

We choose $T_0= 0.18$\,GeV in the parametrization of the bulk viscosity such that the centrality dependence of the identified particles' mean $p_T$ in Pb+Pb collisions at 5.02 TeV can be reproduced. Ref.~\cite{Schenke:2020mbo} used a lower $T_0 = 0.16$\,GeV, which was obtained by matching to results from RHIC measurements.
The larger $T_0$ leads to a weaker centrality dependence of $\langle p_T\rangle$, agreeing better with the ALICE data. We introduce a temperature-dependent $\eta/s$ below $T_0$ in this work to get a better description of the $v_n\{2\}$ measurements beyond the 40\% centrality bin. We find a constant $\eta/s$ would overestimate $v_n\{2\}$ compared to the ALICE measurements.

For every collision system and parameter set, we simulate at least 100k IP-Glasma + hydrodynamics minimum bias events with an average of 100 oversampled hadronic events in the UrQMD phase. These high statistics simulations ensure small enough statistical errors on the observable ratios we present in the following section, allowing us to quantify the impact of the nuclear structure.

\section{Probing nuclear structure in heavy-ion collisions}

\subsection{Imaging the deformation of \Xe{} at the LHC}

The observables in relativistic heavy-ion collisions usually have a complex dependence on model parameters. In this section, we will first identify observables that are insensitive to the model parameters not related to nuclear structure, such as sub-nucleonic structure and the QGP viscosities. Such exploration is essential to find observables that will maximize the sensitivity to the nuclear structure of the colliding nuclei.

\begin{figure}[t!]
    \centering
    \includegraphics[width=\linewidth]{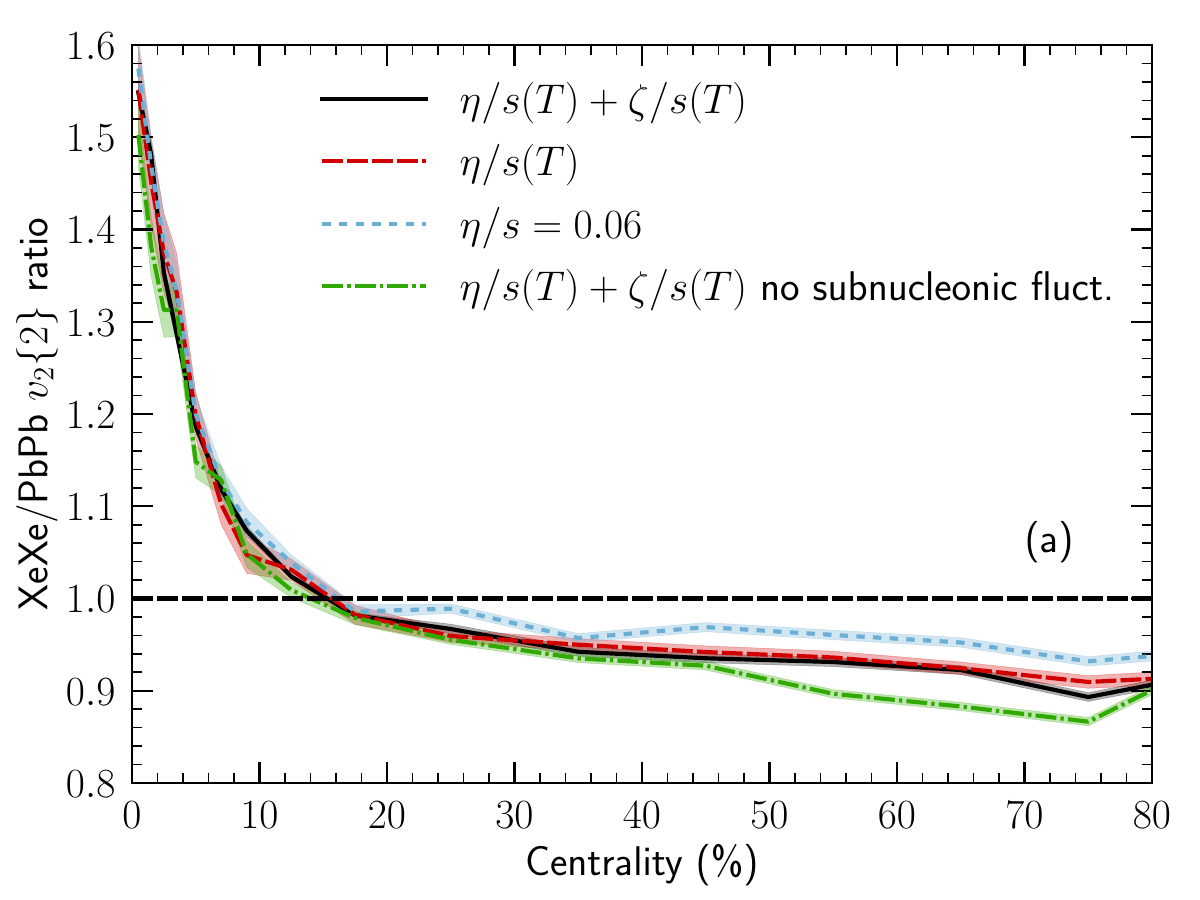}
    \includegraphics[width=\linewidth]{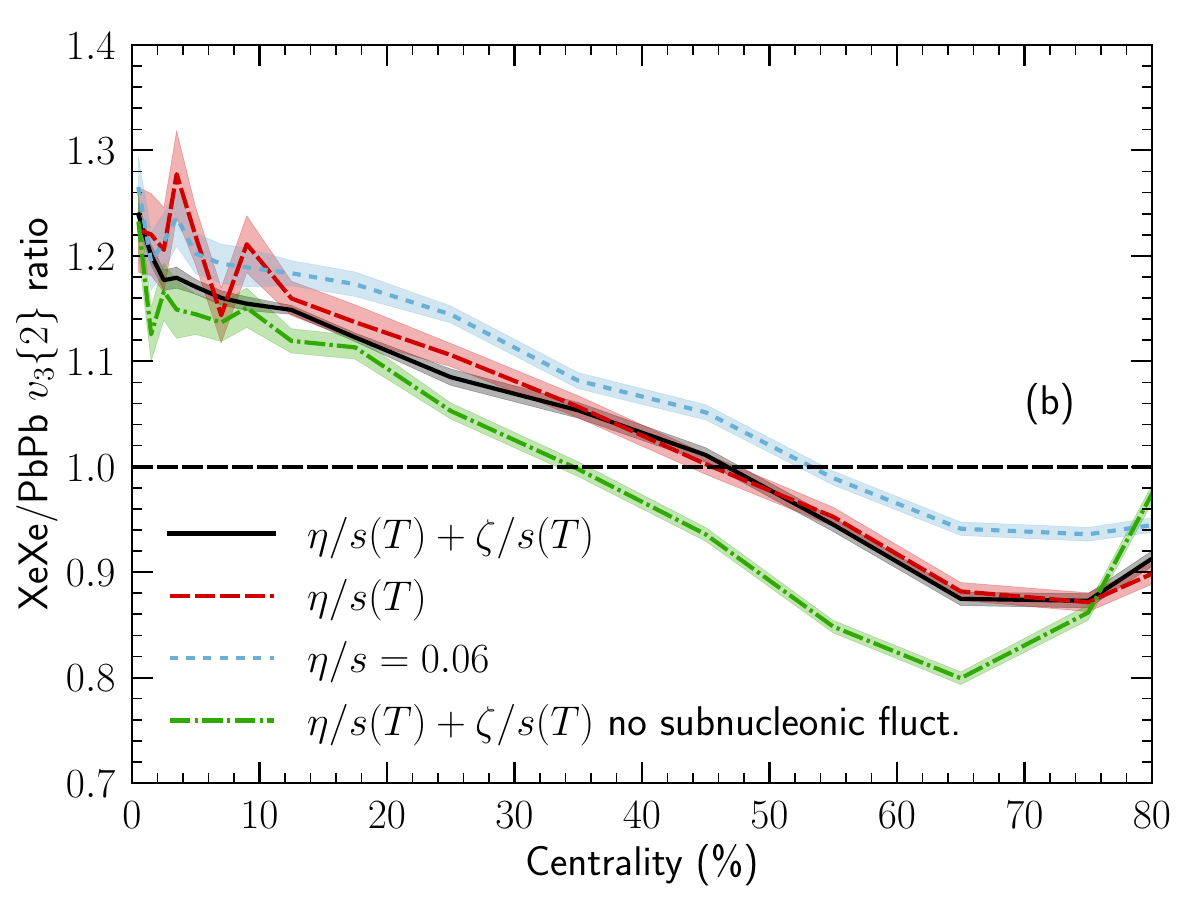}
    \caption{The elliptic (a) and triangular (b) flow ratios between \Xe+\Xe{} collisions at $\snn = 5.44$ TeV and \Pb+\Pb{} collisions at $\snn = 5.02$ TeV with different simulation settings. For nuclear structure parameters, we use \Xe(1) and \Pb(default) in Table~\ref{tab:WoodsSaxon}.}
    \label{fig:XeXevsPbPb_modelvar}
\end{figure}

\begin{figure}[th!]
    \centering
    \includegraphics[width=\linewidth]{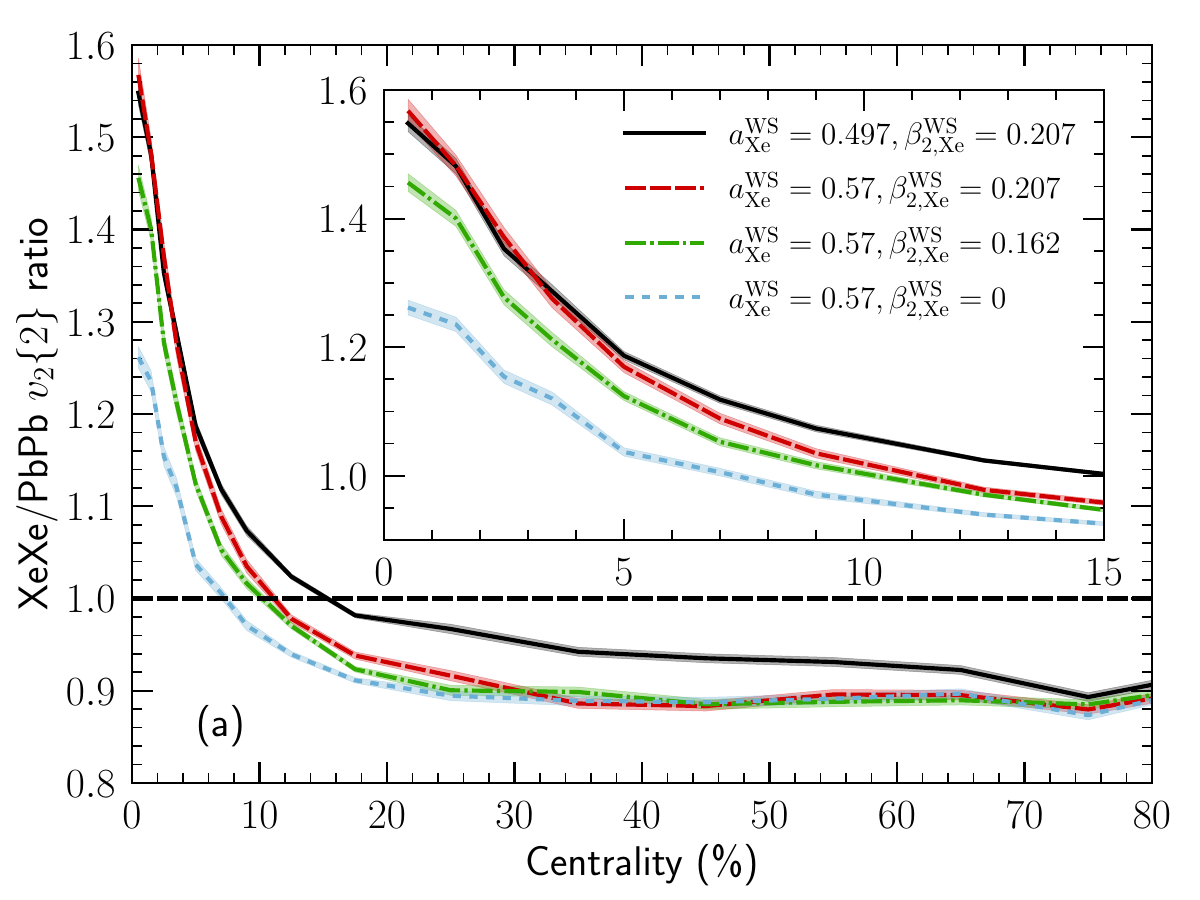}
    \includegraphics[width=\linewidth]{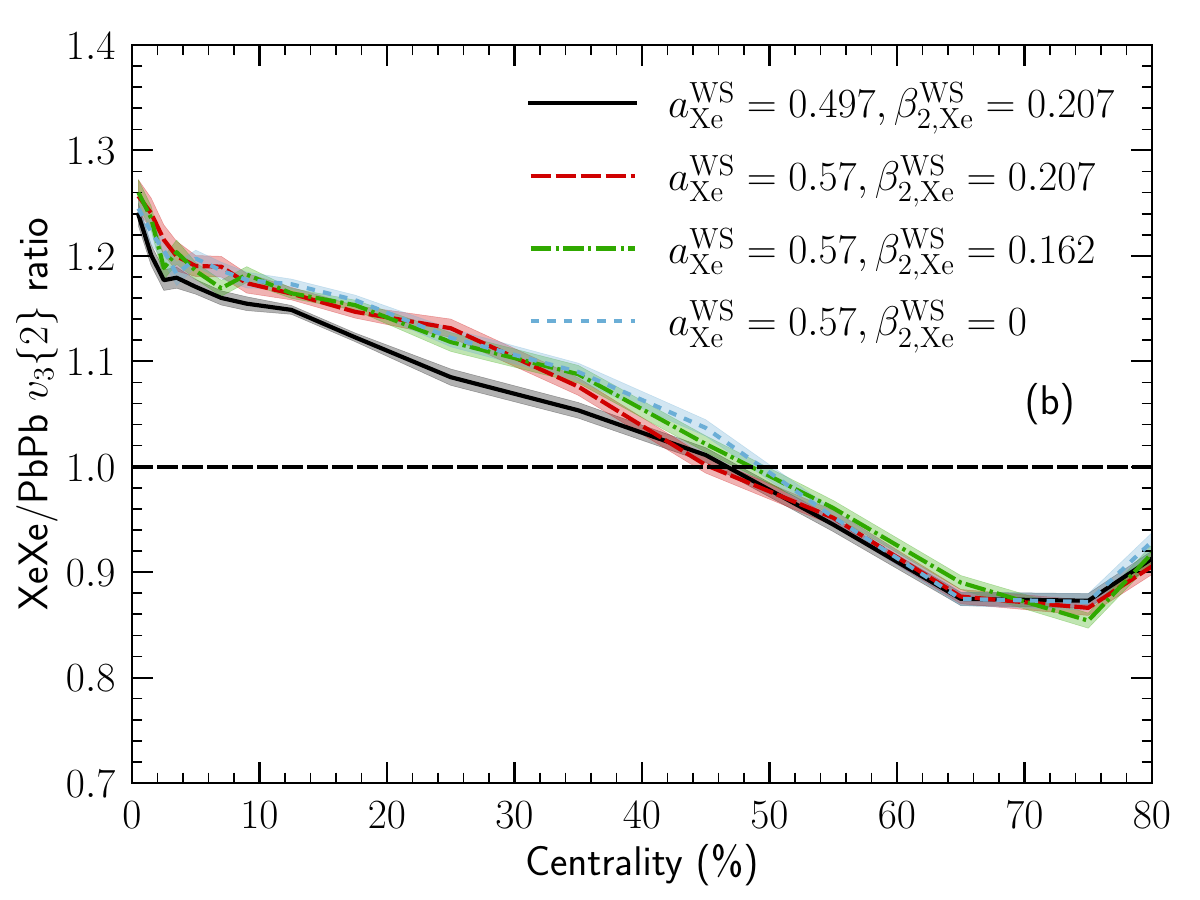}
    \caption{The ratios of $v_n\{2\} (n = 2 (a), 3  (b))$ between \Xe+\Xe{} and \Pb+\Pb{} collisions with different Woods-Saxon parameters for the \Xe{} nucleus in Table~\ref{tab:WoodsSaxon}. The insert in panel (a) zooms in on  the most central events where the effects are largest.}
    \label{fig:XeXevsPbPb_vn2}
\end{figure}

Figure~\ref{fig:XeXevsPbPb_modelvar} shows the ratios of anisotropic flow coefficients $v_2\{2\}$ and $v_3\{2\}$  between \Xe+\Xe{} collisions at $\snn = 5.44$ TeV and \Pb+\Pb{} collisions at $\snn = 5.02$ TeV with four different parameter sets. We find that the elliptic flow ratios between the two collision systems are largely insensitive to the QGP viscosity used in the hydrodynamic simulations. Despite the transverse overlap areas in \Xe+\Xe{} collisions being approximately 25\% smaller than those in \Pb+\Pb{} collisions at the same centrality, the final-state interactions in the hydrodynamics + hadronic transport phases are canceled to very high precision in the ratios of anisotropic flow between two collision systems across all centrality bins. This result demonstrates that experimental measurements of this observable can be used to probe features related to the initial state of heavy-ion collisions.

The IP-Glasma initial-state model includes multiple length scale fluctuations from nuclear and sub-nucleonic structures. To further disentangle their effects in the identified observables, we perform additional simulations with the full shear and bulk viscosity but without sub-nucleonic fluctuations. Figure~\ref{fig:XeXevsPbPb_modelvar} shows that the elliptic flow $v_2\{2\}$ ratios up to 40\% in centrality are insensitive to the sub-nucleonic structures in the initial state model. 

For triangular flow $v_3\{2\}$ ratios, the QGP's specific shear viscosity and sub-nucleonic fluctuations show sizable effects for centralities larger than the 20\% centrality class because the triangular flow is more sensitive to the shorter length scale fluctuations than elliptic flow.
The specific shear viscosity also slightly affects the ratio of flow coefficients, particularly at mid-central and peripheral collisions because of the system size difference between \Xe+\Xe{} and \Pb+\Pb{} collisions.
We have checked that the centrality bin window in which $v_4\{2\}$ and $v_5\{2\}$ ratios are insensitive to the sub-nucleonic fluctuations further shrinks to 0-10\%.

Figure~\ref{fig:XeXevsPbPb_vn2} shows the ratios of anisotropic flow coefficients for different shapes of \Xe{} nuclei with respect to \Pb+\Pb{} collisions. Within the 0--10\% central bins, the values of elliptic flow ratios are sensitive to the elliptical deformation $\beta_2^\mathrm{WS}$ of the \Xe{} nuclei. In 0--1\% centrality, the $v_2\{2\}$ ratios increase by about 20\% as we change $\beta^\mathrm{WS}_{2,\mathrm{Xe}}$ from 0 to 0.207. This result is in line with previous studies comparing Pb+Pb and Xe+Xe collisions~\cite{Giacalone:2017dud} and others that studied the deformation of $^{238}$U nuclei with respect to the measurements in $^{197}$Au+$^{197}$Au collisions at the top RHIC energy~\cite{Ryssens:2023fkv, Fortier:2024yxs}.

The elliptic flow $v_2\{2\}$ ratios in semi-peripheral centralities (20--40\%) exhibit sizable dependencies on the nuclear skin thickness parameter $a^\mathrm{WS}$. This occurs because a smaller $a^\mathrm{WS}$ leads to a sharper edge in the nuclear density profile, which impacts \Pb{} and \Xe{} nuclei differently due to their distinct average radii. The recent ALICE measurements of this ratio \cite{ALICE:2024nqd} could thus serve as a sensitive probe to discern differences in the nuclear skin thickness between the \Xe{} and \Pb{} nuclei.

Figure~\ref{fig:XeXevsPbPb_vn2}b shows that the effects of the \Xe{} nucleus' $\beta_2^\mathrm{WS}$ on $v_3\{2\}$ ratios are negligible in central collisions. We find that a smaller nuclear skin thickness for the \Xe{} nucleus results in a smaller $v_3\{2\}$ ratio around the 20--30\% centrality bin. But sub-nucleonic fluctuations also have noticeable effects on this ratio around this centrality bin as shown in Fig.~\ref{fig:XeXevsPbPb_modelvar}. Combining the ratios of $v_2\{2\}$ and $v_3\{2\}$ could help us disentangle effects of sub-nucleonic fluctuations and skin thickness.

Similar weak dependence on $\beta^\mathrm{WS}_2$ of the \Xe{} nucleus is found for ratios of high-order anisotropic flow coefficients $v_4\{2\}$ and $v_5\{2\}$. However, the initial-state elliptic deformation of the nuclei can affect high-order anisotropic flow vectors ($Q_4$ and $Q_5$) via the nonlinear response effects during hydrodynamic evolution~\cite{McDonald:2016vlt,Zhao:2017yhj,Giacalone:2018wpp, Jia:2022qrq}. 
Such non-linear response to initial geometry is one of the signatures that can demonstrate that the underlying system is strongly coupled. Therefore, it is worthwhile to quantitatively explore the effects of elliptic deformation in central \Xe+\Xe{} collisions with different $\beta^\mathrm{WS}_2$ on both the $v_2\{2\}$ ratio, sensitive to linear response, and the ratios of high-order flow correlations, that are sensitive to non-linear response effects.
Here, we analyze the ratios of the following Pearson coefficients for $Q_4$ and $Q_5$ between the two collision systems, 
\begin{align}
    \rho_{422} &= \frac{\Re\{\langle Q_4 (Q_2^2)^* \rangle\} }{\sqrt{\langle |Q_4|^2 \rangle \langle |Q_2|^4 \rangle}} \,,\\
    \rho_{523} &= \frac{\Re\{\langle Q_5 (Q_2 Q_3)^* \rangle\} }{\sqrt{\langle |Q_5|^2 \rangle \langle |Q_2|^2 |Q_3|^2 \rangle}}\,.
\end{align}
Here, $Q_n \equiv \sum_j \exp(i n \phi_j)$ is the $n$-th order complex anisotropic flow vector and $\Re\{\cdots\}$ takes the real part of the correlation function.

\begin{figure}[t!]
    \centering
    \includegraphics[width=\linewidth]{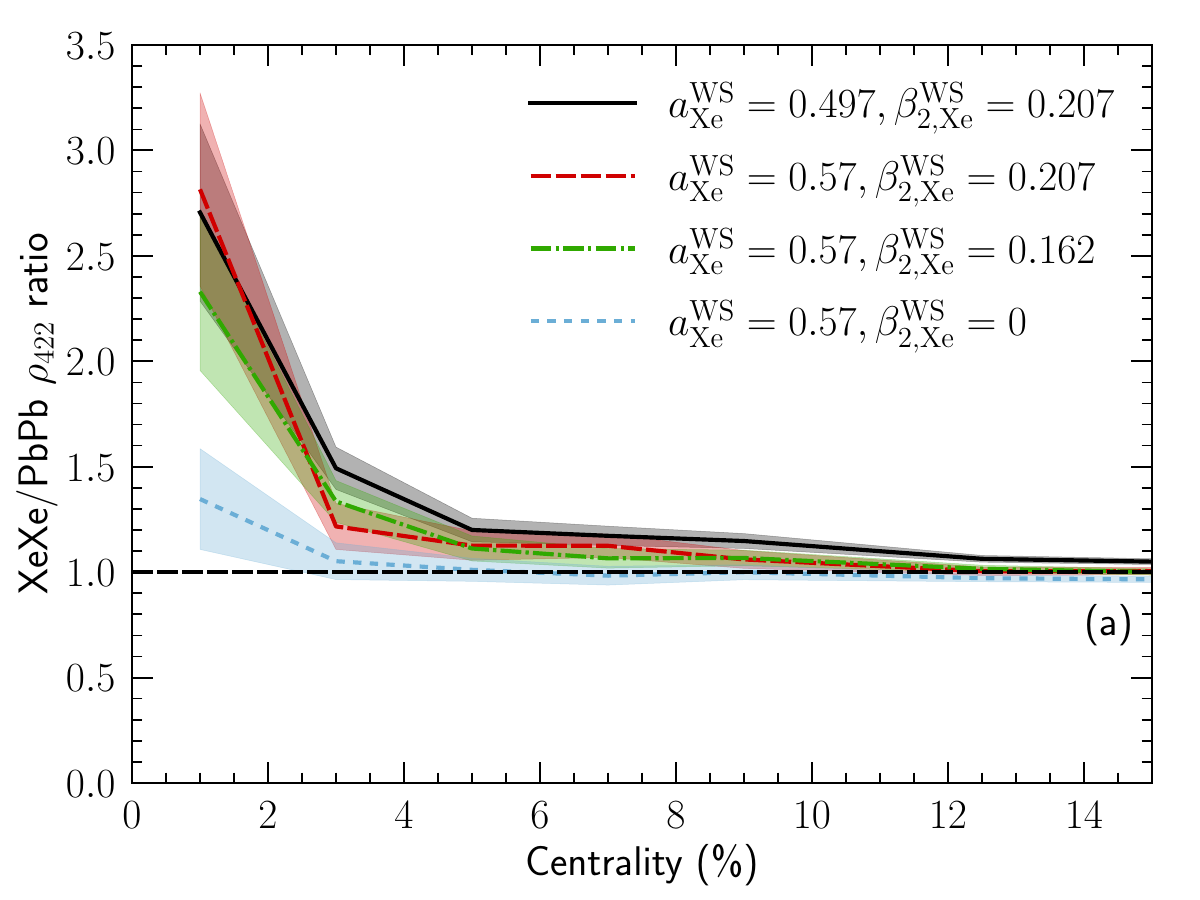}
    \includegraphics[width=\linewidth]{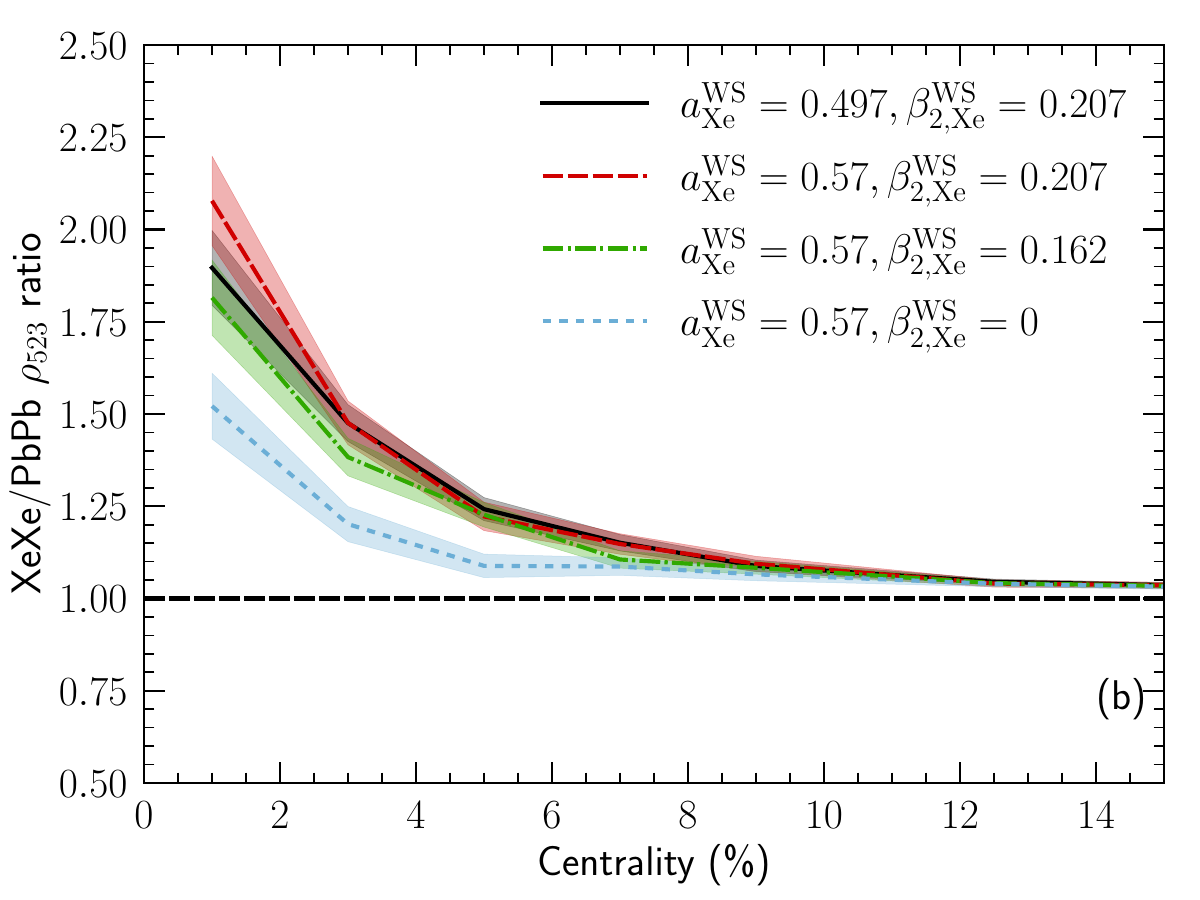}
    \caption{The ratios of non-linear correlation coefficients $\rho_{422}$ (Panel (a)) and $\rho_{523}$ (Panel (b)) between \Xe+\Xe{} and \Pb+\Pb{} collisions with different Woods-Saxon parameters for the \Xe{} nucleus in Table~\ref{tab:WoodsSaxon}.}
    \label{fig:XeXevsPbPb_v45Ratio}
\end{figure}

Figure~\ref{fig:XeXevsPbPb_v45Ratio} presents the ratios of the non-linear response coefficients $\rho_{422}$ and $\rho_{523}$ between \Xe+\Xe{} and \Pb+\Pb{} collisions. Our results indicate that the larger elliptical deformation of the \Xe{} nucleus leads to increased $\rho_{422}$ and $\rho_{523}$ coefficients in central collisions. The sensitivity of these coefficients to the deformation parameter $\beta^\mathrm{WS}_{2,\mathrm{Xe}}$ is comparable to that observed for the $v_2\{2\}$ ratios. We expect these measurements in central collisions to provide complementary constraints to $v_2\{2\}$ measurements on the $\beta_2^\mathrm{WS}$ deformation of the colliding nucleus. The recent ALICE measurement \cite{ALICE:2024nqd} of the $\rho_{422}$ ratio between these two systems reaches a value above 1.7 for central collisions (0-5\%), consistent with our model calculations where $\beta^\mathrm{WS}_{2,\mathrm{Xe}} > 0$.

\begin{figure}[t!]
    \centering
    \includegraphics[width=\linewidth]{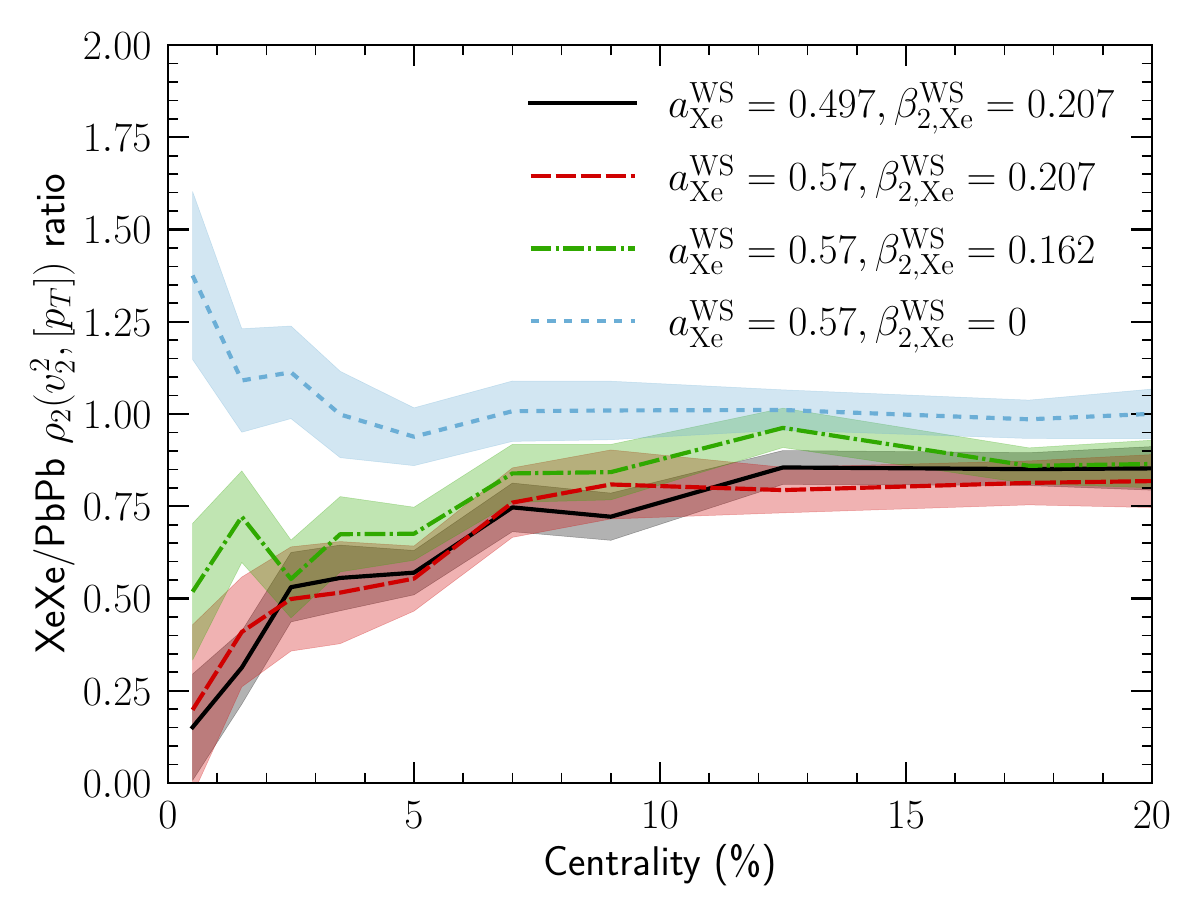}
    \caption{The ratios of $v_2-[p_T]$ correlations between \Xe+\Xe{} and \Pb+\Pb{} collisions with different Woods-Saxon parameters for the \Xe{} nucleus.}
    \label{fig:XeXevsPbPb_rhonRatio}
\end{figure}

The observable $v_2-[p_T]$ correlation is sensitive to the non-trivial correlation between the system's elliptical deformation and system size. It is defined as
\begin{align}
    \rho_2 (v_2^2, [p_T]) = \frac{\langle |Q_2|^2 ([p_T] - \langle [p_T] \rangle) \rangle}{\sqrt{\langle |Q_2|^4 \rangle \langle ([p_T] - \langle [p_T] \rangle)^2 \rangle}},
\end{align}
where $[p_T]$ denotes the total transverse momentum of all charged hadrons within one collision event and $\langle [p_T] \rangle$ is the averaged total transverse momentum over a centrality bin class.

Figure~\ref{fig:XeXevsPbPb_rhonRatio} presents the ratios of the $v_2-[p_T]$ correlations between \Xe+\Xe{} and \Pb+\Pb{} collisions. For central collisions, the \Xe{} nucleus with a larger $\beta_2^\mathrm{WS}$ leads to a smaller ratio. This is because, when nuclei are deformed, the elliptic flow in central collisions comes from both the deformed geometry and event-by-event shape fluctuations. Generally, shape fluctuations significantly contribute to the $v_2-[p_T]$ correlation. A large $\beta_2^\mathrm{WS}$ deformation in the colliding nucleus diminishes the relative contribution of shape fluctuations to the elliptic flow, resulting in a weaker $v_2-[p_T]$ correlation. Varying $\beta^\mathrm{WS}_{2, \mathrm{Xe}}$ between 0 and 0.207, the $\rho_2$ ratio varies from 1.5 to 0.25 in central collisions, showing strong sensitivity to the elliptical deformation of the \Xe{} nucleus~\cite{Bally:2021qys}. The nuclear surface thickness $a^\mathrm{WS}_\mathrm{Xe}$ does not have a significant effect on the ratios.

\begin{figure}[h!]
    \centering
    \includegraphics[width=\linewidth]{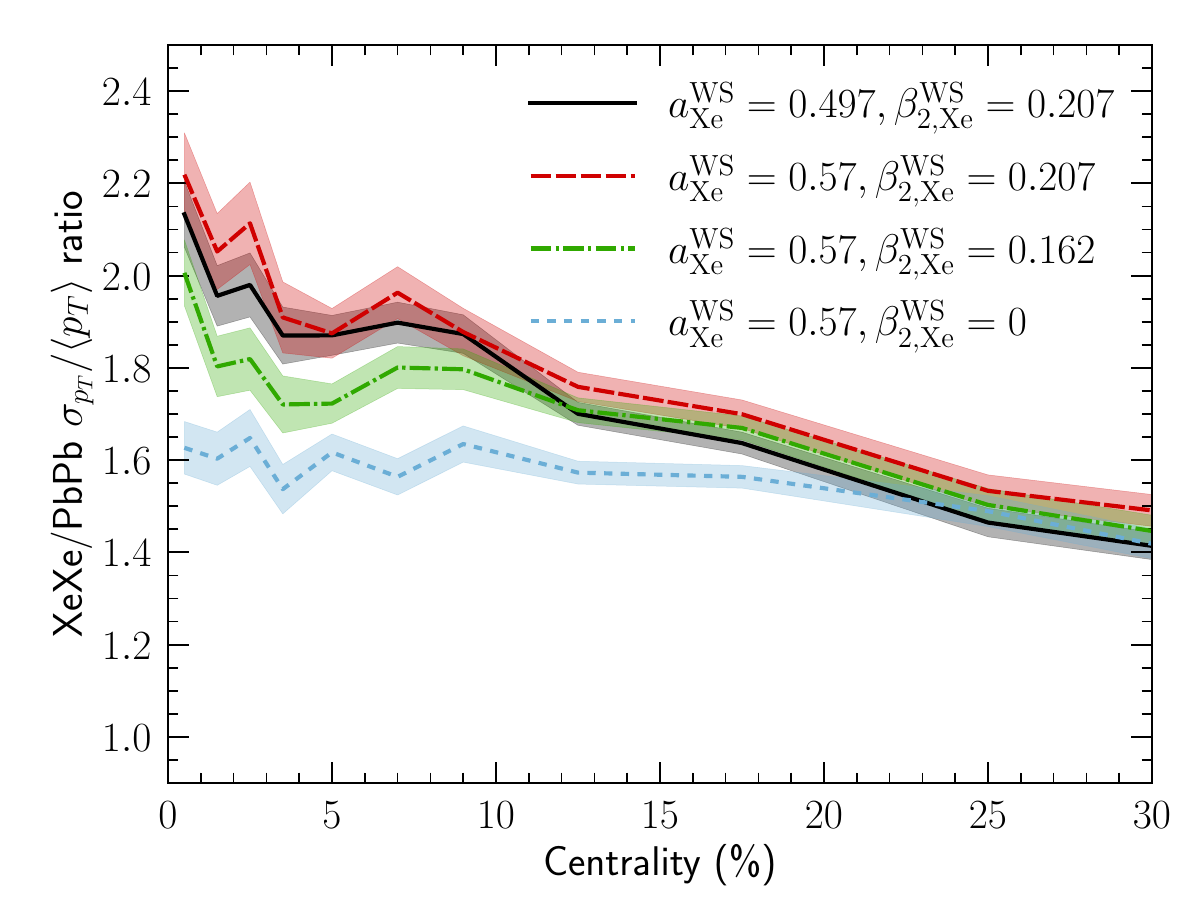}
    \caption{The ratio of the normalized variance of charged hadron transverse momentum fluctuations, $\sigma_{p_T}/\langle p_T \rangle \equiv \sqrt{\langle (p_T - \langle p_T \rangle)^2 \rangle}/\langle p_T \rangle$, between \Xe+\Xe{} and \Pb+\Pb{} collisions with different Woods-Saxon parameters for the \Xe{} nucleus.}
    \label{fig:pTVarRatio}
\end{figure}

Figure~\ref{fig:pTVarRatio} shows the ratio of the normalized variance of charged hadron transverse momentum between the two collision systems.
The ratio is overall above unity because \Xe{} nuclei have fewer nucleons than \Pb{} nuclei and therefore contain more shape and size fluctuations which lead to a larger variance in the final-state hadrons' transverse momenta. 
We find the \Xe{} nuclei with larger elliptical deformation to result in a larger variance of transverse momentum fluctuations in central collisions. This result is intuitive to understand as a non-zero elliptic deformation introduces more shape and size fluctuations in the collisions~\cite{Dimri:2023wup}.

\subsection{Constraining the neutron skin of \Pb{}}

Another interesting topic in nuclear structure studies is the neutron skin size of the \Pb{} nucleus. Although the majority of observables studied in high-energy heavy-ion collisions do not explicitly depend on the difference between protons and neutrons in the initial state, we will explore some observables that are sensitive to the overall nuclear skin thickness of the colliding nuclei. These observables provide indirect information for constraining the neutron skin if the charge radius of the \Pb{} nucleus can be accessed from complementary low-energy experiments and its evolution to high energy is estimated.

\begin{figure}[t!]
    \centering
    \includegraphics[width=\linewidth]{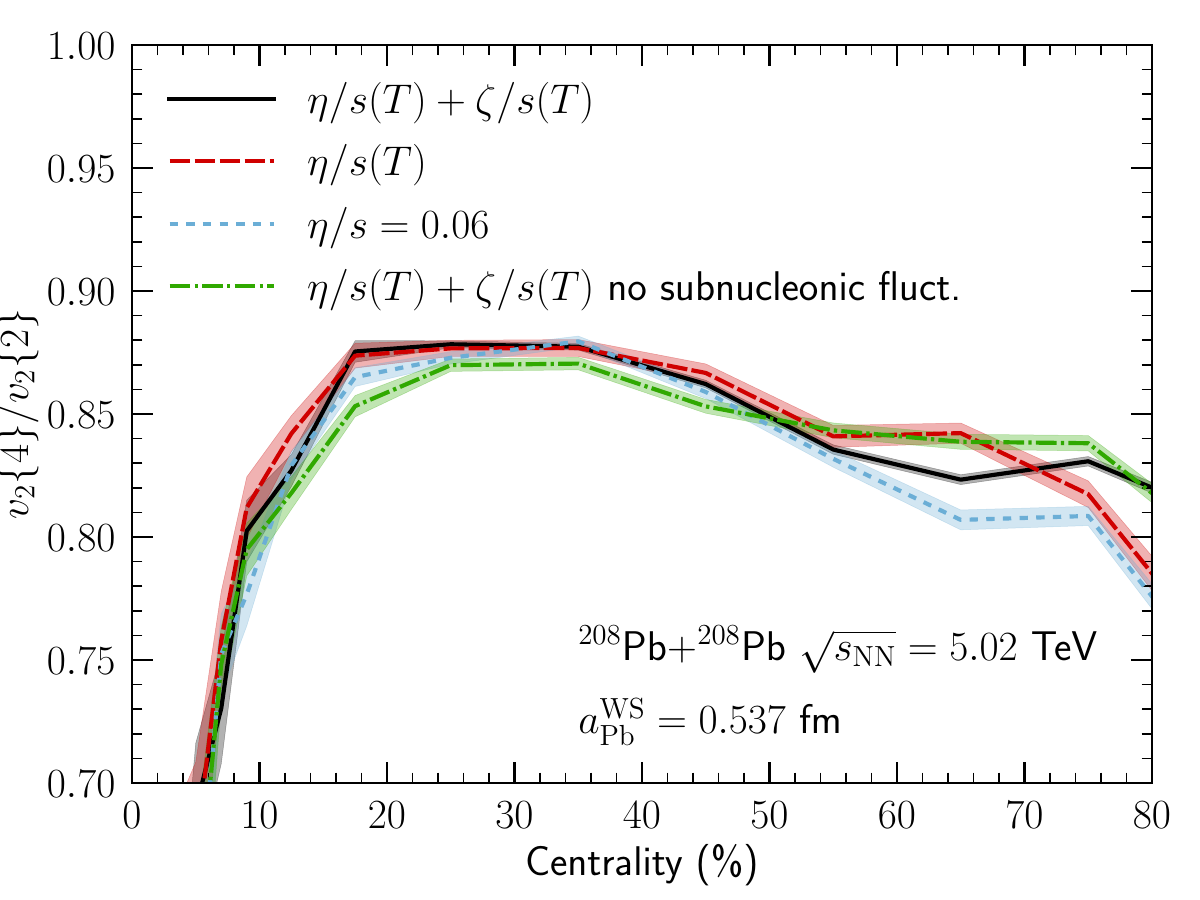}
    \caption{The ratios of charged hadron $v_2\{4\}$ and $v_2\{2\}$ in \Pb+\Pb{} collisions at $\snn = 5.02$ TeV using the \Pb(default) in Table~\ref{tab:WoodsSaxon} and different simulation settings.
    }
    \label{fig:PbPbv2nRatioParamDep}
\end{figure}

We find one promising observable, the $v_2\{4\}/v_2\{2\}$ ratio, which is sensitive to the nuclear skin thickness of the colliding nuclei. Ref.~\cite{Jia:2022qgl} studied related observables based on the TRENTO \cite{Moreland:2014oya}  initial-state model for the RHIC isobar collisions.  The $v_2\{4\}/v_2\{2\}$ ratio does not require measurements in two different collision systems. Figure~\ref{fig:PbPbv2nRatioParamDep} shows that the $v_2\{4\}/v_2\{2\}$ ratios in \Pb+\Pb{} collisions are insensitive to the choices of QGP viscosity and initial-state sub-nucleonic fluctuations over a wide range of centrality.

We use the following Woods-Saxon parameters for the \Pb{} nucleus $R_{0,p}^\mathrm{WS} = 6.68$\,fm, $a^\mathrm{WS}_p = 0.448$\,fm, and $\beta_2^\mathrm{WS} = 0.006$. 
 These parameters lead to a Root-Mean-Square (RMS) radius for the proton density of $R_p = 5.435$\,fm. We introduce different sizes of neutron skins by setting $R_{0,n}^\mathrm{WS} = 6.69$\,fm with different neutron skin thicknesses $a^\mathrm{WS}_n$ listed in Table~\ref{tab:WoodsSaxonPbNeutronSk}~\cite{Trzcinska:2001sy, Zenihiro:2010zz, Giacalone:2023cet}. One can compute the neutron skin of the \Pb{} nucleus using the RMS radii difference between proton and neutron densities,
\begin{align}
    \Delta R_{np} &= R_p - R_n, \\
    R_{p,n} &= \sqrt{\int \mathrm{d}^3 r r^2 \rho^\mathrm{WS}_{p,n}(r, \theta) \bigg/ \int \mathrm{d}^3 r \rho^\mathrm{WS}_{p,n}(r, \theta)}.
\end{align}

\begin{table}[tb]
    \centering
    \caption{The values for the Woods-Saxon parameters for $^{208}$Pb with different sizes of neutron skin $\Delta R_{np}$ used in the IP-Glasma initial conditions~\cite{Trzcinska:2001sy, Zenihiro:2010zz}.}
    \begin{tabular}{c|c|c|c}
        \hline \hline 
        Nucleus & $R_{0,n}^\mathrm{WS} - R_{0,p}^\mathrm{WS}$ & $a^\mathrm{WS}_n - a^\mathrm{WS}_p$ (fm) & $\Delta R_{np}$ (fm) \\ \hline
        \Pb(1) & 0.01 & 0.119 & 0.149 \\ \hline
        \Pb(2) & 0.01 & 0.160 & 0.201 \\ \hline
        \Pb(3) & 0.01 & 0.198 & 0.250 \\ \hline
        \hline
    \end{tabular}
    \label{tab:WoodsSaxonPbNeutronSk}
\end{table}

\begin{figure}
    \centering
    \includegraphics[width=\linewidth]{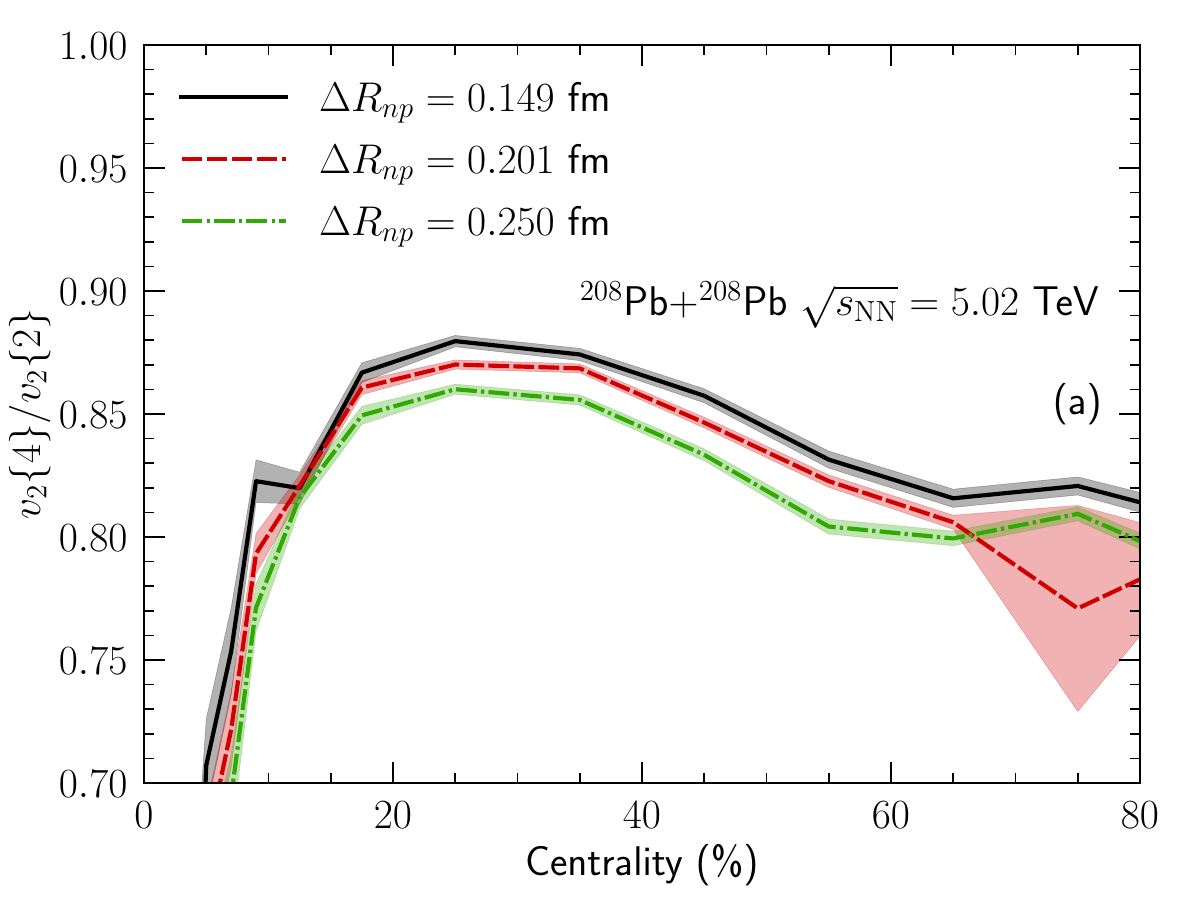}
    \includegraphics[width=\linewidth]{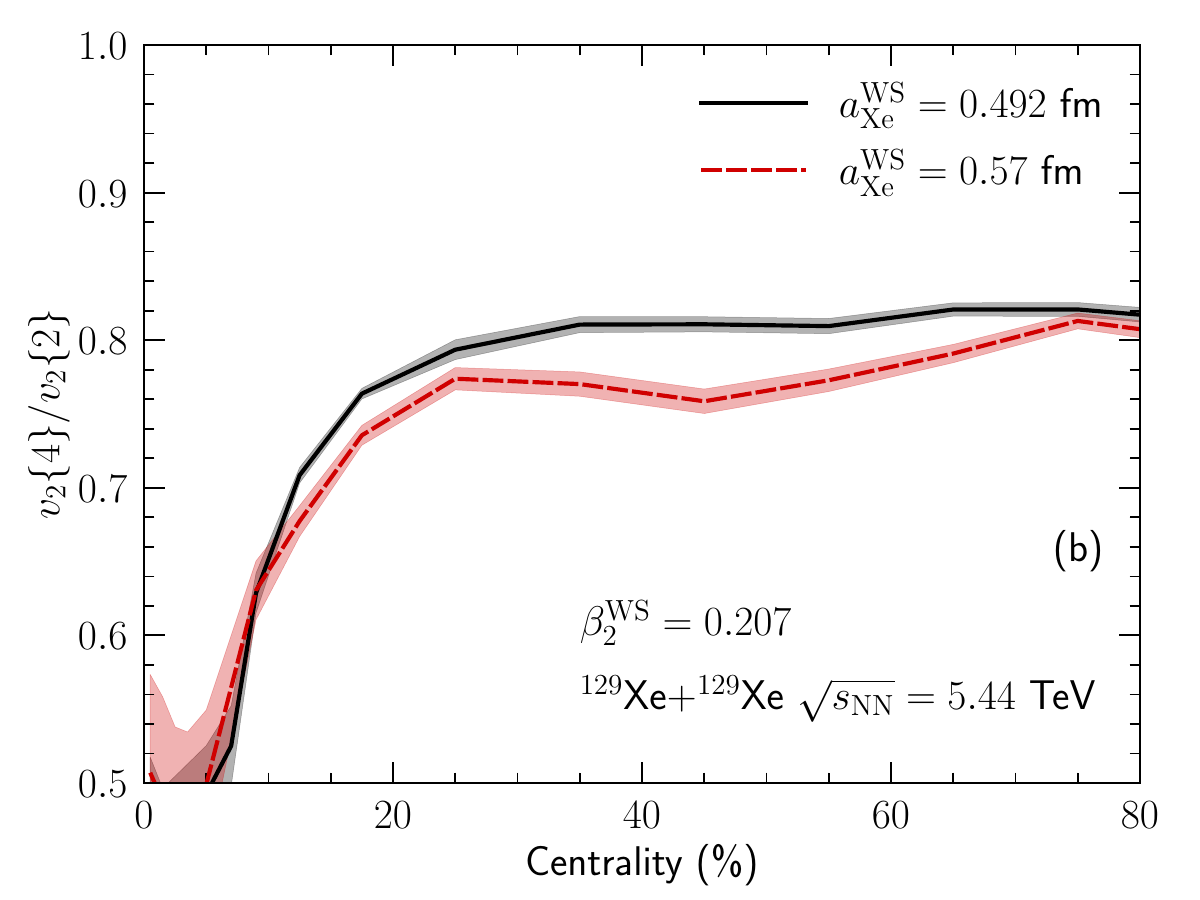}
    \caption{The ratios of $v_2\{4\}$ to $v_2\{2\}$ as functions of centrality for \Pb+\Pb{} collisions with different sizes of neutron skin (Panel (a)) and \Xe+\Xe{} collisions with different nuclear surface thickness (Panel (b)).}
    \label{fig:v2nRatio}
\end{figure}

Figure~\ref{fig:v2nRatio}a shows the effects of varying the \Pb{} neutron skin on the $v_2\{4\}/v_2\{2\}$ ratio at high energies. A smaller neutron skin results in a larger $v_2\{4\}/v_2\{2\}$ ratio in 20-60\% semi-peripheral collisions.
Assuming that the elliptic flow has a Bessel-Gaussian distribution with mean $\bar{v}_2$ and variance $\sigma_{v_2}^2$~\cite{Ollitrault:2009ie}, then
\begin{align}
    \frac{v_2\{4\}}{v_2\{2\}} = \sqrt{\frac{\bar{v}^2_2 - \sigma_{v_2}^2}{\bar{v}^2_2 + \sigma_{v_2}^2}} = \sqrt{\frac{1 - \tilde{\sigma}_{v_2}^2}{1 + \tilde{\sigma}_{v_2}^2}},
\end{align}
where $\tilde{\sigma}_{v_2}^2 \equiv \sigma_{v_2}^2/\bar{v}_2^2$ is the normalized variance of elliptic flow. A decreasing $v_2\{4\}/v_2\{2\}$ ratio with increasing $\Delta R_{np}$ indicates that a larger nuclear skin depth generates more fluctuations in elliptic flow. This is because a larger nuclear skin depth allows nucleons to be more diffusively populated around the edge of the overlapping area, increasing the shape fluctuations.

We also compute this ratio observable for \Xe+\Xe{} collisions and show its dependence on the nuclear skin thickness $a^\mathrm{WS}$ in Fig.~\ref{fig:v2nRatio}b. The results convey the same physics message as for \Pb+\Pb{} collisions.

Future precision measurements of the $v_2\{4\}/v_2\{2\}$ ratio can serve as a sensitive probe for the skin depth of the nuclear mass density of the colliding nuclei. By combining the knowledge of charge radius from low-energy nuclear experiments, one can deduce the size of the neutron skin of the colliding nuclei.

\section{Conclusion}

In this work, we identified several experimental observables, which show strong sensitivity to the nuclear structure inputs in the initial state of high-energy heavy-ion collisions. Although there is a 40\% difference in the mass numbers between the \Xe{} and \Pb{} nuclei, we find that final-state effects from hydrodynamics and hadronic transport are canceled to very high precision in the ratios of anisotropic flow coefficients between the two systems, in particular for the most central collisions. Precision measurements of these two systems at the LHC  can provide valuable information for constraining the shape deformation of the \Xe{} nucleus. The ratios of non-linear response coefficients $\rho_{422}$ and $\rho_{523}$ between the two collision systems are also sensitive to the elliptical deformation of the \Xe{} nucleus. The experimental measurements~\cite{ALICE:2024nqd} can provide complementary information to constrain the value of $\beta^\mathrm{WS}_{2,\mathrm{Xe}}$.

We also find that the nuclear skin thickness is sensitive to the elliptic flow fluctuations which result in measurable effects of varying the skin thickness on the ratio of $v_2\{4\}/v_2\{2\}$ in semi-peripheral collisions.

High statistics simulations as performed in this work are essential for constraining the nuclear structure information in phenomenological studies of heavy-ion collisions. They are needed for quantitative model-to-data comparisons to access information about the event-by-event shape fluctuations of the colliding nuclei. Our simulation results serve as a benchmark for comparisons with recent and upcoming measurements at the Large Hadron Collider.

Our event-by-event simulation data are open-source and can be used by community users \cite{shen_2024_13839384}.

\section*{Acknowledgements}
We thank Y. Zhou for requesting high-statistics calculations for measurements from the ALICE Collaboration, which motivated this work. We also thank G. Giacalone for useful discussions.
This material is based upon work supported by the U.S. Department of Energy, Office of Science, Office of Nuclear Physics, under DOE Contract No.~DE-SC0012704 (B.P.S.) and Award No.~DE-SC0021969 (C.S.), and within the framework of the Saturated Glue (SURGE) Topical Theory Collaboration.
C.S. acknowledges a DOE Office of Science Early Career Award. 
H.M. is supported by the Research Council of Finland, the Centre of Excellence in Quark Matter, and projects 338263, 346567, and 359902, and by the European Research Council (ERC, grant agreements No. ERC-2023-101123801 GlueSatLight and ERC-2018-ADG-835105 YoctoLHC).
W.B.Z. is supported by DOE under Contract No. DE-AC02-05CH11231, by NSF under Grant No. OAC-2004571 within the X-SCAPE Collaboration, and within the framework of the SURGE Topical Theory Collaboration. 
The content of this article does not reflect the official opinion of the European Union and responsibility for the information and views expressed therein lies entirely with the authors.
This research was done using resources provided by the Open Science Grid (OSG)~\cite{Pordes:2007zzb, Sfiligoi:2009cct, OSG1, OSG2}, which is supported by the National Science Foundation award \#2030508 and \#1836650.

\bibliography{inspirehep, non-inspirehep}

\end{document}